# A Review of Augmented Reality Applications for Building Evacuation

Ruggiero Lovreglio

Lecturer, School of Engineering and Advanced Technology, Massey University, Auckland, New Zealand.
Email: r.lovreglio@massey.ac.nz, website: www.lovreglio.info

**Abstract:**

Evacuation is one of the main disaster management solutions to reduce the impact of man-made and natural threats on building occupants. To date, several modern technologies and gamification concepts, e.g. immersive virtual reality and serious games, have been used to enhance building evacuation preparedness and effectiveness. Those tools have been used both to investigate human behavior during building emergencies and to train building occupants on how to cope with building evacuations.

Augmented Reality (AR) is novel technology that can enhance this process providing building occupants with virtual contents to improve their evacuation performance. This work aims at reviewing existing AR applications developed for building evacuation. This review identifies the disasters and types of building those tools have been applied for. Moreover, the application goals, hardware and evacuation stages affected by AR are also investigated in the review. Finally, this review aims at identifying the challenges to face for further development of AR evacuation tools.

## 1. INTRODUCTION

Building evacuations are one of the key risk reduction strategies for buildings affected by man-made and natural disasters. To date, many studies have been carried out focusing on the investigation of human behavior in disasters such as fire and earthquakes and to simulate those behavior in evacuation model tools (Gwynne *et al.*, 1999; Kuligowski, Peacock and Hoskins, 2010; Lindell and Perry, 2012; Lovreglio *et al.*, 2017).

Novel technologies, such as virtual reality and serious games, have been proved to be fundamental to investigate human behavior in disasters and to enhance building occupants' preparedness. For instance, virtual reality has been used (1) to investigate the occupant perception of evacuation systems (Ronchi and Nilsson, 2015; Olander *et al.*, 2017); (2) to investigate evacuee route and exit choices (Andrée, Kristin, Daniel Nilsson, 2015; Cosma, Ronchi and Nilsson, 2016; Lovreglio, Fonzone and Dell'Olio, 2016); (3) evacuee navigation (Lovreglio, Ronchi and Nilsson, 2015b; Moussaïd *et al.*, 2016; Kyriakou, Pan and Chrysanthou, 2017; Rios, Mateu and Pelechano, 2108); and to train building occupants for fire and earthquake emergencies as reviewed in (Lovreglio *et al.*, 2018; Zhenan Feng *et al.*, 2018).

Augmented Reality (AR) is a novel technology which is gaining popularity among the public since the release of new AR ardware (i.e. Microsoft HoloLens, Meta 2) and software in the last years (i.e. Pokemon Go). Furthermore, AR is a technology that could help enhance building evacuation performance by providing users with real-time digital information. This would allow users to make real-time in-place best actions decisions during evacuations and emergencies.

The aim of this work is to review existing AR applications for building evacuation to identify which fields and disasters those tools have been applied to and to highlight the challenges to face while developing an AR application for building evacuation.

## 2. BACKGROUND

### 2.2 Augmented Reality

Augmented Reality (AR) is a live view of the real-world environment whose elements are "augmented" by computer-generated visual information (Carmigniani *et al.*, 2011). As such, this technology occupies a specific place in the reality-virtuality spectrum illustrated in Figure 1.

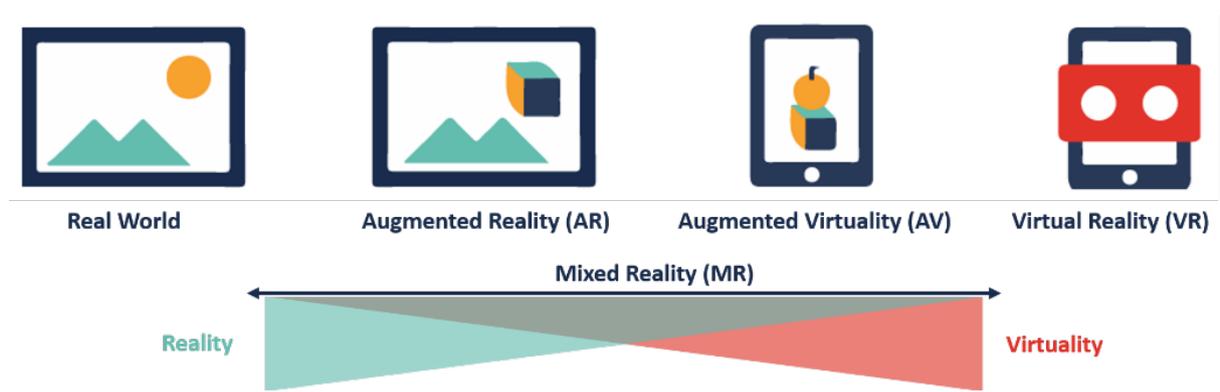

Figure 1 – Reality-virtuality spectrum

From Figure 1, it is possible to observe that AR occupies the left-hand side of the reality-virtuality spectrum as the main component is reality while the computer-generated visual information is a secondary component augmenting the reality.

AR can provide both a direct or indirect live view of the augmented environment. A direct view can be provided using lenses where computer-generated information is projected to be reflected in the user's eyes while an indirect view is obtained by using a camera and a display showing live and augmented video stream (Carmigniani *et al.*, 2011).

These AR algorithms usually consist of two stages: tracking and reconstructing/recognizing. The tracking can be done using markers (e.g. QR codes) or using markerless systems. In the first case, the algorithms constantly search for defined markers to impose digital elements using those markers as reference points. In the second case, more advanced techniques, such as Simultaneous Localization And Mapping (SLAM) and Structure from Motion (SfM) are used for mapping fiducial markers relative positions (Carmigniani *et al.*, 2011). A review of AR tracking, interaction and display solutions is available in (Feng Zhou, Duh and Billinghurst, 2008; Wagner and Schmalstieg, 2009).

**2.1 Building Evacuation**

To investigate AR applications for building evacuation, it is worth briefly describing the stage and events characterizing evacuations. The evacuation of a building threatened by a man-made or natural disasters can be seen as the sum of several sequential events. The time required to complete a building evacuation is known in the literature as the required safe egress time (RSET). This time is required to be smaller than the available safe egress time (ASET), which is the time available to evacuate a building safely before the building conditions become untenable for occupants (Lovreglio, 2016). RSET can be divided into several subsequetial times as shown in Figure 2. As such, several steps need to be completed to have a successful evacuation.

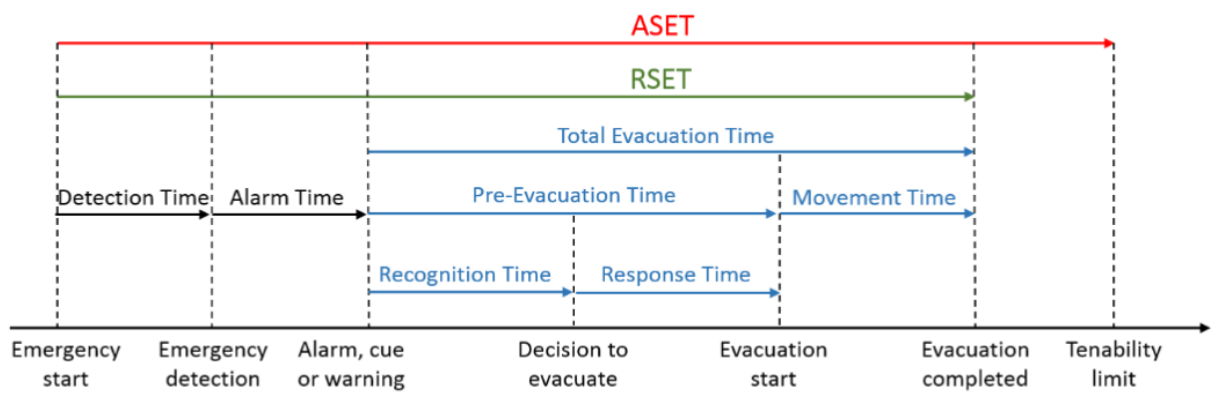

Figure 2 – Timeline building evacuation framework (Lovreglio, 2016)

One of the most critical time is the total evacuation time since it accounts for human behaviour and response to an emergency. For some types of disasters, such as fire and bushfire, the pre-evacuation stage can be the most critical one as building occupants need time to recognise the emergencies and to get ready before the actual evacuation (Lovreglio, Ronchi and Nilsson, 2015a, 2016). For other disasters, such as earthquakes, building occupants can easily perceive that the building is shaking while they might require time to respond (Lovreglio *et al.*, 2017).

The final stage of a building evacuation, it is the actual movement towards a safe place, e.g. an assembly point. In this stage, building occupants are asked to find a safe way out using the available zinformation provided by fire wardens and building evacuation systems such as evacuation lightings.

Virtual reality and augmented virtuality have been broadly used to investigate human behavior and to train building occupants in different evaluation stages. To date, several reviews are available in the literature investigating different evacuation purposes of virtual reality tools see for instance (Kinateder *et al.*, 2014; Zhenan Feng *et al.*, 2018). However, a comprehensive literature review of AR applications for building evacuation is still not available. As such, this research aims at covering this research gap.

## 2. METHOD

### 2.1 Review Objective

The proposed review of AR applications for building evacuations is designed to fulfil the following three objectives:

1. The identification of the building disasters where the AR applications have been applied;
2. The investigation which evacuation stage has been affected by the AR application;
3. The study of the hardware used for AR applications.

Those objectives have been defining to have a general overview of this subject as well as identify research gaps for future research and AR development.

### 2.2 Work Selection

The papers selected for this review were collected from journals, conferences, patents and reports. The papers were recovered from the following databases: Google Scholar and Scopus.

The selected works needed to address two major concepts: augmented reality and building evacuation. The search was conducted using the following combination of key words: *augmented reality* or *smart glasses* + *evacuation* or *building evacuation*. Only works written in English were included in the first search. Such a search was run on the 1$^{st}$ of April 2018 and yield to the first selection of 22 works out of 200 works accessed through Scopus and Google Scholar. This first selection was done making sure that the selected key words were at least mentioned in the title and abstract of those works.

A second selection was then carried out considering the two selection criteria:
(a) An AR application for building evacuation was proposed;
(b) An AR application was tested through experiments.

Only 12 papers met at least of the two criteria and were finally included in the review. The full selection process described in this section is illustrated in Figure 3.

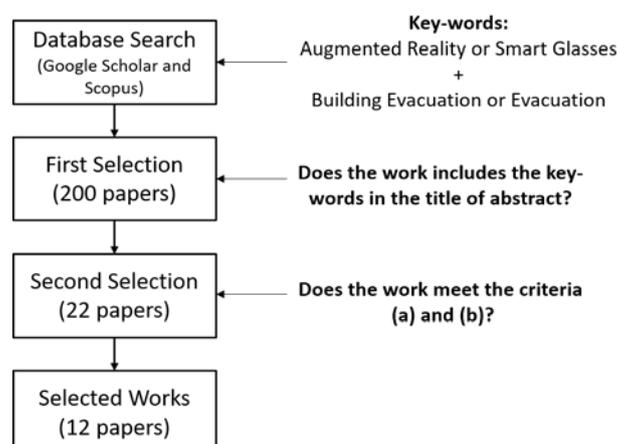

Figure 3 – Selection process for the review.

## 3. RESULTS

A list of the selected papers is provided in Table 1. The results indicate that all the selected papers have been published in the last eight years (the least recent work was published in 2010) showing that applications of AR for building evacuation are recent research goals. Besides the general applications listed in Table 1, most of the remaining applications focus on Tsunami, Earthquake and Fire building evacuations. Moreover, the results indicate that those applications have been mainly developed for educational buildings and large-scale buildings.

### 3.1 Application Goals

This review identified three types of AR application for building evacuation.

The majority of the applications (5 out of 9) were developed for *training purposes*. Those applications were conceived to overcome the well known limitations of traditional evacuation drills which are reviewed in (Gwynne *et al.*, 2016, 2017). As such, those applications trying to enhance the realism and engagement of the evacuation drills showing the impact of the threats on the buildings and its occupants. For instance, Kawai et al. (2016a) López et al. (2010) augmented drills with digital fires, smoke, injured occupants and cracks generated by an earthquake and building fire. Iguchi et al. (2016) and Mitsuhara et al. (2017) also augmented their earthquake drills with the damage generated as well as with digital building occupants the participants are supposed to interact with.

Another type of AR application goal is to *navigate building occupants* during building evacuations. Those applications were developed for large-scale buildings having complex geometries to help building occupants with wayfinding information. For instance, Ahn and Han (2011) and Cai et al. (2011) developed a navigation system that potentially can be used for any types of disaster while López (2010) proposed a smart system for fire emergencies supporting occupants depending on their characteristics and fire location.

Finally, the latest type of application is to *visualize building evacuation simulations*. This solution was recently developed by Lochhead and Hedley (2018) to investigate if and how AR can be used to evacuee movement simulated by existing evacuation models. As such, they provide a new approach to link evacuation simulations with the real-world context of the built environment.

### 3.2 Type of Hardware

Most of the AR applications listed in Table 1 were developed using indirect AR non-immersive and immersive technologies. Smartphones and tablets are the common non-immersive solutions while Oculus and Google cardboards were used as immersive solutions. Although the immersive solutions can enhance the level of realism, Kawai et al. (2016b) reported that they could also produce motion sickness and discomfort for users.

In few instances, direct AR technologies were used. For example, Kawai et al. (2016a) and Mitsuhara et al. (2017) tested smart glasses named Moverio. Both references argued that the main limitations of Moverio are the small field of view where digital contents can be visualized and the user difficulty in interacting with the device.

To date, there is only one study comparing pros and cons of different AR pieces of hardware from a user viewpoint. Mitsuhara et al. (2017) tested and compared the use of Oculus, Google cardboard and Moverio for training purpose. After testing those three alternatives, the participants rated those solutions regarding usability and visual capabilities. The results indicate that Google cardboards had the best performance.

Table 1 shows that many AR application running on smartphones and tables required markers as those devices are not capable of handling the SLAM and SfM algorithms described in Section 2.1. In some occasions, those devices are used without markers as there is a cloud system locating the devices and providing them with the appropriate digital contents.

### 3.3. Evacuation Stages

All the reviewed applications focus on the movement stage during building evacuations. In fact, all the navigation systems in Table 1 are developed to provide wayfinding information to building occupants to enhance their evacuation performance. On the other hand, the training applications used digital contents to enhance the realism of the evacuation routes with the cues of disasters (e.g. fire and smoke) or the impact of the disasters on the building and its occupants (e.g. cracks and injured occupants).

Table 1 – Selected papers for the review

| Reference | Disaster | Goal | Type of hardware | Marker based | Type of Building | Evacuation Stage |
|---|---|---|---|---|---|---|
| (Kawai, Mitsuhara and Shishibori, 2016a) (Mitsuhara *et al.*, 2016) | Tsunami | Training | Moverio - smart glasses (direct AR) | No | University | Movement |
| (Kawai, Mitsuhara and Shishibori, 2016b) (Kawai, Mitsuhara and Shishibori, 2015) | Tsunami Earthquake | Training | Oculus HMD (indirect AR) Tablet based (indirect AR) | Yes | High Schools | Movement (visualization of earthquake impact) |
| (Ahn and Han, 2011) (Ahn and Han, 2012) | General | Indoor navigation | Smartphone (indirect AR) | No | Large scale buildings | Movement |
| (Mitsuhara, Iguchi and Shishibori, 2017) | Earthquake | Training | Oculus HMD (indirect AR) Cardboard smartphone (indirect AR) Moverio - smart glasses (direct AR) | Yes only for Cardboard and Mover | Pre schools | Pre-evacuation and Movement |
| (Iguchi, Mitsuhara and Shishibori, 2016) | Earthquake | Training | cardboard smartphone (indirect AR) | Yes | Pre schools | Pre-evacuation and Movement |
| (Ortakci *et al.*, 2017) | Fire | Indoor navigation | Smartphone (indirect AR) | No | Large scale buildings | Movement |
| (López *et al.*, 2010) | Fire | Training | Smartphone (indirect AR) | No | General | Movement (enhancing realism) |
| (Lochhead and Hedley, 2018) | General | Simulation Visualization | Smartphone and Tablet (indirect AR) | Yes | General | Movement |
| (Yi-Wen Cai and Shih-Cheng Wang, 2011) | General | Indoor navigation | Smartphone and Tablet (indirect AR) | No | General | Movement |

Only a couple of applications also focus on the pre-evacuation stage. Mitsuhara et al. (2017) and Iguchi et al. (2016) developed a training AR application for school teachers who are asked to interact with digital pre school students during an earthquake shaking before starting evacuating after the end of the shake.

## 4. DISCUSSION AND CONCLUSION

This paper introduces a literature review of the existing Augmented Reality (AR) applications developed for building evacuations. This review identifies 11 relevant conference and journal papers and 1 USA patent.

This review shows that several applications have been developed mainly for disasters such as tsunamis, earthquakes and fires affecting educational buildings and large-scale buildings. As such, applications for many other man-made or natural disasters (e.g. terroristic attacks, bushfires and hurricanes) have not been investigated yet.

The reviewed works indicate that AR applications have been used for training purpose to enhance the realism of traditional evacuation drills adding digital contents. Moreover, several applications have been developed with the aim of providing building occupant with wayfinding solutions to enhance their evacuation performance. A novel application is the use AR to visualize the result of evacuation simulation in real-world contexts.

This paper shows that several indirect and direct AR technologies have been used. The review indicates that indirect immersive AR solutions such as Oculus can generate motion sickness while direct AR solutions such as Moverio have a limited field of view that reduces the immersion. It is possible to argue that those existing limitations can be easily overcome by using more recent direct AR hardware solutions such as Microsoft HoloLens and Meta 2. However, implementations using those two innovative solutions are still not available in the literature.

This review illustrates that all the AR applications were developed to affect the movement stage during building evacuations and only in two instances those applications were used for training purpose during the pre-evacuation stage. Therefore, future studies are necessary to investigate how AR tools can enhance pre-evacuation training and how those tools can support building occupants during real emergencies providing them with the best action to take.

### ACKNOWLEDGEMENTS
Ruggiero would like to thank Prof Robyn Phipps for letting him start his research and teaching journey at Massey University (NZ) and Lara Tooky and Phil Jackson for proof reading this work.